\documentclass[fleqn,usenatbib,useAMS]{rasti}

\usepackage{graphicx}
\usepackage{amsmath}
\usepackage{amssymb}
\usepackage{booktabs}
\usepackage{float}
\usepackage{caption}
\usepackage{hyperref}
\usepackage{orcidlink}
\usepackage{xcolor} 
\usepackage{soul}
\usepackage{lastpage}
\graphicspath{{figures/}
}
\definecolor{lightblue}{RGB}{210,235,255}
\sethlcolor{lightblue}
\soulregister\citep7
\soulregister\citet7
\soulregister\ref7

\title[HERMES]{HERMES: HiERarchical Modelling
       for Exoplanet Science}

\author[W.\,M.\,F.\ Naqvi \& N.\,B.\ Cowan]{%
Wasi M.\ F.\ Naqvi\orcidlink{0009-0001-9895-7593}$^{1}$ and
Nicolas B.\ Cowan\orcidlink{0000-0001-6129-5699}$^{1,2}$\\
$^{1}$Department of Physics, McGill University, 3600 rue University, Montréal QC H3A 2T8, Canada\\
$^{2}$Department of Earth \& Planetary Sciences, McGill University, 3450 rue University,
Montréal, QC H3A 0E8, Canada\\}

\pubyear{{\the\year{}}}

\begin{document}
\label{firstpage}
\pagerange{\pageref{firstpage}--\pageref{LastPage}}
\maketitle

\begin{abstract}
ESA's Ariel Space Mission will characterise the atmospheres of
          $\sim\!1000$ exoplanets to quantify population-level trends. We present HERMES (HiERarchical Modelling for Exoplanet
          Science), a multidimensional Bayesian framework that probes
          population-level correlations across multiple axes of diversity. 
          The specific use case we present is the multidimensional relation between stellar metallicity, planetary mass, and atmospheric metallicity.
Starting from the Ariel Mission Candidate Sample, we select confirmed planets with available masses and stellar metallicity, inject plausible multidimensional trends and
          demonstrate successful parameter recovery.
 Simulated surveys are generated with a variety of
          leverage and sample size, in the presence of  intrinsic astrophysical scatter and measurement noise.
 By fitting independent Bayesian models to each survey, we confirm that survey leverage remains a reliable predictor of trend
          precision even in multiple dimensions and in the presence of intrinsic astrophysical scatter. 
          For an Ariel Tier 2 transit survey of at least 400 planets, HERMES robustly recovers the correlation between stellar and planetary metallicity despite intrinsic scatter in planetary abundances as large as 1.2 dex. These results
  establish HERMES as a practical tool for survey design and science yield
  forecasting in preparation for Ariel and other surveys probing population-level trends.
\end{abstract}


\section{Introduction}\label{sec:intro}

Exoplanet science has entered an era of large-scale comparative atmospheric
  characterisation. Where early studies demonstrated that individual planetary
  atmospheres could be chemically probed through transmission and emission
  spectroscopy \citep{Charbonneau2002,Charbonneau_2005}, the frontier is shifting to
  understanding
  these atmospheres \emph{statistically}, seeking the population-level trends
  that encode how planets form and evolve. This shift builds on earlier population-level atmospheric work, from the statistical
interpretation of Spitzer thermal phase variations by \cite{Cowan_2011} to the comparative Hubble
transmission-spectral survey of hot Jupiters by \cite{Sing_2016}. 

ESA's Ariel mission will be the
  first space-based facility purpose-built for this goal: over a four-year primary mission
  it
  will deliver 0.55--7.8\,$\mu$m transmission and emission spectra for approximately 1000
  exoplanets \citep{Tinetti2018,tinetti2021}. This
  will provide an ensemble for rigorous empirical tests of physical trends across the exoplanet
  population. Among the most fundamental of these trends is the relationship between
  atmospheric metallicity and planetary mass. In the solar system, the four
  giant
  planets follow a clear inverse pattern: more massive planets are less enriched
  in heavy elements relative to hydrogen, with metallicities ranging from
  $\sim$3--6 times solar for Jupiter to $\sim$70--100 times solar for Uranus and
  Neptune \citep{Guillot2005}. This is broadly understood within the
  core-accretion paradigm, in which more massive planets accrete proportionally
  more hydrogen and helium during runaway gas accretion. A central open question in planetary science is whether this inverse trend persists across the
  diverse exoplanet population, spanning hot Jupiters, warm Neptunes, and
  sub-Neptunes in a wide range of orbits.

Progress on this question has come primarily from transmission spectroscopy of
  close-in exoplanets, whose high temperatures keep water in the gas phase and
  accessible to remote sensing \citep[e.g.][]{Madhusudhan2011,Kreidberg2014}.  \citet{Kreidberg2014} provided one of the first
  precise water abundance measurements for an exoplanet, finding the
  $2\,M_\mathrm{J}$ hot Jupiter WASP-43b has a water content of 0.4--3.5
  times solar at $1\sigma$ confidence, a metallicity consistent with an
  extrapolation of the solar system's inverse trend to higher mass. Building on
  such individual benchmarks, \citet{welbanks2019} used the Hubble and Spitzer Space Telescopes for a
  homogeneous population-level retrieval study, measuring H$_2$O, Na, and K
  abundances for 19 transiting exoplanets spanning cool mini-Neptunes to hot 
  Jupiters. More recently, JWST has sharpened these constraints: the compilation of \citet{Lothringer2026} assembles atmospheric abundances across the observed sample and finds a robust mass--metallicity trend, in fact somewhat \emph{steeper} than that of the Solar System giants and interior-structure predictions, demonstrating that population-level trends are now detectable with space-based telescopes.
  
  Whether the mass--metallicity relation reflects true bulk metallicity or
  is
heavily influenced by high carbon-to-oxygen ratios, clouds, or formation history
  remains
  actively debated \citep{IkomaKobayashi2025}, and more recent reanalyses with
  updated stellar abundance measurements have found the statistical evidence for
   a
  mass slope to be weaker than originally reported \citep{Sun2024}, highlighting
  the sensitivity of current inferences to sample size and data homogeneity. Population-synthesis modelling by \citet{Fortney2013} first made this link explicit, predicting that planetesimal accretion enriches the H/He envelope to metal mass fractions of $\sim$100--400$\times$ solar for low-mass planets, and establishing the theoretical basis for connecting a planet's bulk heavy-element content to its observable atmospheric metallicity. A
  complementary dimension is the role of host-star
  metallicity: because protoplanetary disk composition broadly tracks stellar
  composition, one expects the reservoir of heavy elements available for
  planetary
  accretion to scale with $[\mathrm{Fe/H}]_\star$, leaving a potentially
  detectable imprint in planetary atmospheric abundances.  
  
  Even at fixed planetary mass and stellar abundance, atmospheric metallicities are not necessarily
  deterministic: planets may differ in accretion history, migration
  pathway, envelope loss, cloud opacity, atmospheric mixing, and elemental
  abundance ratios. This intrinsic astrophysical scatter can broaden the observed
  mass--metallicity relation beyond measurement uncertainty alone
  \citep{Swain_2024}. Disentangling a
  stellar
  metallicity signal from the planetary mass trend and from intrinsic
  astrophysical scatter demands both the large samples that Ariel will provide
  and
  statistical tools capable of handling them along multiple axes of diversity 
  simultaneously.

  Hierarchical Bayesian frameworks are the natural tool for
  population-level inference: jointly
  inferring individual planetary properties and the hyperparameters
  that describe the underlying trend while propagating all sources of
  uncertainty \citep{Keating2022,Lustig_Yaeger_2022}. A model is \emph{hierarchical} when it is structured in levels: individual objects are described by unobserved properties, and those properties are in turn governed by shared population-level parameters, or hyperparameters. It is \emph{Bayesian} when every parameter is assigned a prior and inference returns full posterior distributions, so that measurement uncertainty and population-level variance are propagated coherently rather than as point estimates \citep{Eadie2016}. 

  In this paper we present HiERarchical Modelling for Exoplanet
  Science (HERMES), a multidimensional Bayesian hierarchical framework that extends the
  standard 2D mass--metallicity model to three dimensions, simultaneously
  inferring how atmospheric metallicity scales with both planetary mass and
  host-star metallicity. Starting from the Ariel Mission Candidate Sample, we inject physically
  motivated trends, construct mock surveys across a range of sample sizes and
  mass-leverage values, and demonstrate robust parameter recovery. We then
  quantify how increasing intrinsic astrophysical scatter limits the ability of
  the 3D model to detect the stellar--planetary metallicity correlation
  and identify the survey designs for which this distinction remains
  possible. 
  
  The paper is structured as follows. Section~\ref{sec:design}
  describes
  the data, mock survey construction, leverage metrics, and the hierarchical
  model.
  Section~\ref{sec:fits} presents posterior reconstructions and z-score
  calibration. Sections~\ref{sec:expected} and \ref{sec:stellar_det} quantify
  the
  leverage dependence of parameter precision and the recovery of the
  stellar--planetary metallicity correlation, respectively. Section~\ref{sec:summary} summarizes our results and discusses implications for Ariel target selection
  and future surveys.

\section{Methods}\label{sec:design}

\subsection{Data and Survey Design}\label{sec:mcs}

The Ariel Mission Candidate Sample (MCS)
  \citep{Edwards_and_Tinetti}
  is a curated target list of planets suitable for Ariel
  observations,
  selected using signal-to-noise estimates from ArielRad
  \citep{ArielRad}.
  We use the MCS release dated 11 May 2026 \citep[private communication from B. Edwards;][]{MCS_github}. Although
  the full catalogue contains 977 confirmed planets, we
  restrict the
  analysis to systems with available host-star metallicity
  measurements,
  $[\mathrm{Fe/H}]_\star$, and physically valid planetary-mass
  uncertainty
  bounds, yielding a final sample of 858 planets. For each planet, we use three quantities: the logarithmic
  planetary mass,
  $\log(M/M_J)$; the host-star metallicity, $[\mathrm{Fe/
  H}]_\star$, reported
  with asymmetric uncertainties; and the atmospheric Carbon and Oxygen
  abundance (used here as a proxy for planetary atmospheric metallicity),
  $\log [{\mathrm{(C+O)/H}}]$, which is treated as the response
  variable. We assign a fixed symmetric uncertainty of 0.2 dex to
  $\log [{\mathrm{(C+O)/H}}]$ as an idealized high-quality space-based
  retrieval precision. This value is motivated by Tier 2 retrieval forecasts for suitable giant-planet targets \citep{daoust2025}, and is consistent with recent Ariel information-content studies \citep{Radica2026}. An Ariel Tier 2 transit observation comprises of targets selected for higher-precision, deeper spectroscopic survey with a higher signal-to-noise ratio (SNR) to enable more detailed constraints on atmospheric composition. \citep{panek2026,Tinetti2018}

We adopt $(\mathrm{C+O})/\mathrm{H}$ rather than the water abundance alone because
H$_2$O is jointly sensitive to the carbon-to-oxygen ratio, temperature, vertical mixing,
and the sequestration of oxygen into condensates, and is therefore not a one-to-one proxy
for bulk atmospheric metallicity. Ariel's $0.55$--$7.8\,\mu$m coverage constrains the
major carbon carriers (CH$_4$, CO, CO$_2$) alongside H$_2$O, making $(\mathrm{C+O})/\mathrm{H}$
both better motivated physically and closer to what the mission will actually measure
\citep{da_Silva_2024}. We inject a physically motivated mass--metallicity trend, adopting
the slope and intercept measured by \citet{welbanks2019} for water abundances normalised to
their host stars, with an additional host-star metallicity contribution and intrinsic
astrophysical scatter:
  \begin{equation}\label{eq:rel}
  \begin{split}
\log [{\mathrm{(C+O)/H}}]
  &= -1.09\,\log(M/M_J) - 0.95 \\
  &\quad + [\mathrm{Fe/H}]_\star + \mathcal{N}(0,\,0.53).
  \end{split}
  \end{equation}
 Our model assumes that mass-metallicity trends follow a linear relationship such as Equation~\ref{eq:rel}. Recovering this injected relation tests whether the
  hierarchical framework
  can infer the underlying dependence of atmospheric
  metallicity on planetary
  mass and host-star metallicity. Equation~\ref{eq:rel} is an injected test trend rather than an astrophysical claim. It suffices in providing a known ``ground truth'' against which parameter recovery can be measured: more general mass–metallicity relation will flatten towards stellar composition at high mass, and  Equation~\ref{eq:rel}, being a pure power law, does not.

\subsection{Mock Survey Generation: Defining Nested Mass
  Classes}\label{sec:classes}

\begin{figure}
\centering
\includegraphics[width=\columnwidth, alt={Histogram of log planetary mass for nested survey classes S1 through
  S4. Each successive class removes the lowest-mass quartile, narrowing the mass
  range while preserving the high-mass tail.}
]{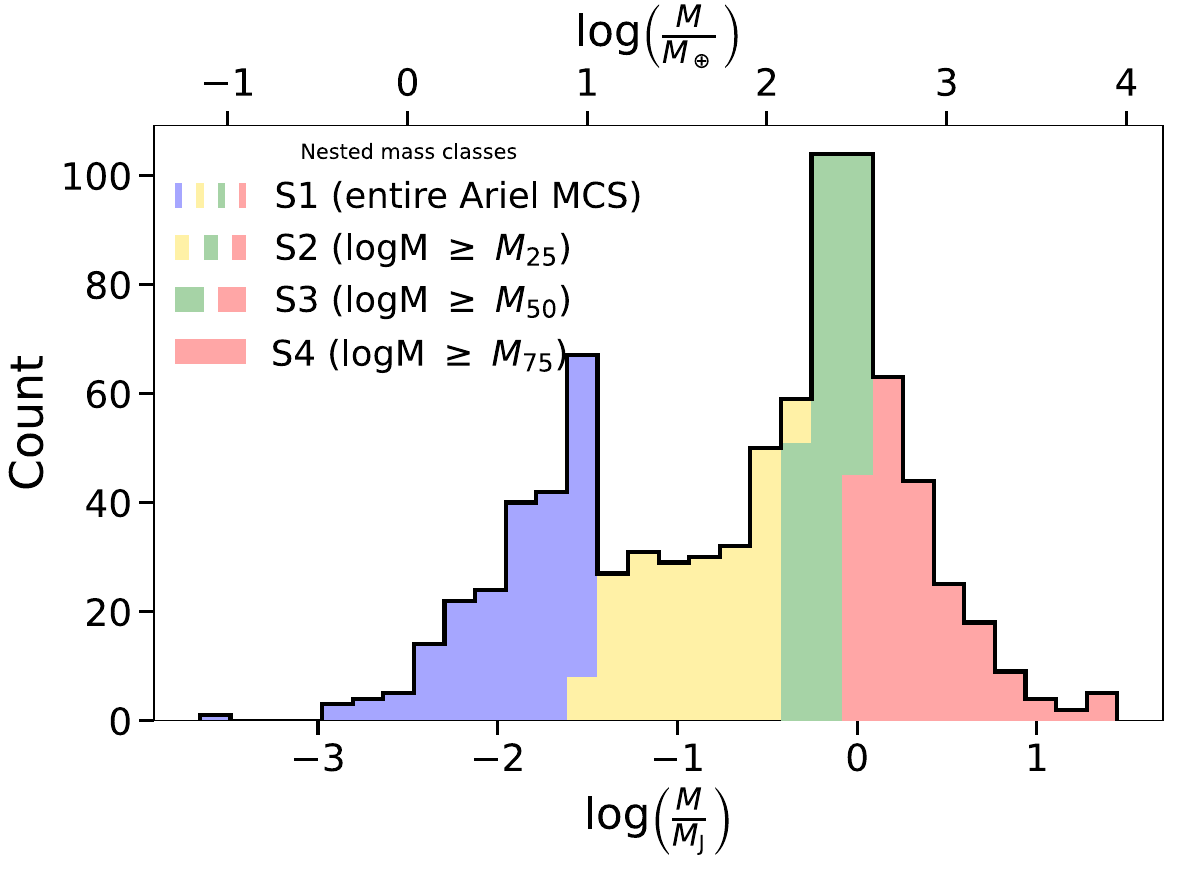}
\caption{Nested mass-class scheme S1--S4. Each successive class
removes the lowest-mass quartile, reducing mass leverage while
preserving the high-mass tail. Surveys are drawn from each class
independently.}
\label{fig:nested}
\end{figure}

To investigate how survey design affects the recovery of the underlying
  mass--metallicity relation, we generated mock surveys using a nested mass-class
scheme based on quartiles of $\log(M/M_J)$ (Figure~\ref{fig:nested} and 
Table~\ref{tab:nested_mass_classes}). Here,
  $M_{25}$, $M_{50}$, and $M_{75}$ denote the 25th, 50th, and 75th percentiles
  of the mass distribution in the filtered MCS. The four nested mass classes, S1 through S4, are constructed by progressively removing planets from the parent sample.  This construction progressively removes the lower-mass portion of the
  sample while retaining the higher-mass tail, providing nested subsamples.

  \begin{center}
  \captionof{table}{Definition of the nested mass classes used to generate mock
  surveys. Quartiles are computed from the full filtered catalogue in
  $\log(M/M_J)$.}
  \label{tab:nested_mass_classes}
  \begin{tabular}{@{}llr@{}}
  \toprule
  Class & Selection criterion & Total Number of planets \\
  \midrule
  S1 & Full catalogue &  858 \\
  S2 & $\log(M/M_J) \geq M_{25}$ &  644 \\
  S3 & $\log(M/M_J) \geq M_{50}$ &  429 \\
  S4 & $\log(M/M_J) \geq M_{75}$ &  215 \\
  \bottomrule
  \end{tabular}
  \end{center}

  For each mass range, mock surveys were drawn at sample sizes
  \[
  N \in \{50, 80, 150, 200, 250, 350, 400, 500, 600\},
  \]
  with five independent realizations generated for every class--size
  combination. All surveys were sampled without replacement from their
  respective parent class, yielding a total of
  $
  4 \times 9 \times 5 = 180
  $
  mock surveys.
  
  \begin{figure}
  \centering
  \includegraphics[width=1\columnwidth,alt={Scatter plot of survey sample size versus planetary-mass dispersion.
  Points are coloured by mass class S1 through S4, with constant-leverage
  contours showing that larger sample sizes and wider mass ranges produce higher
  mass leverage.}]
  {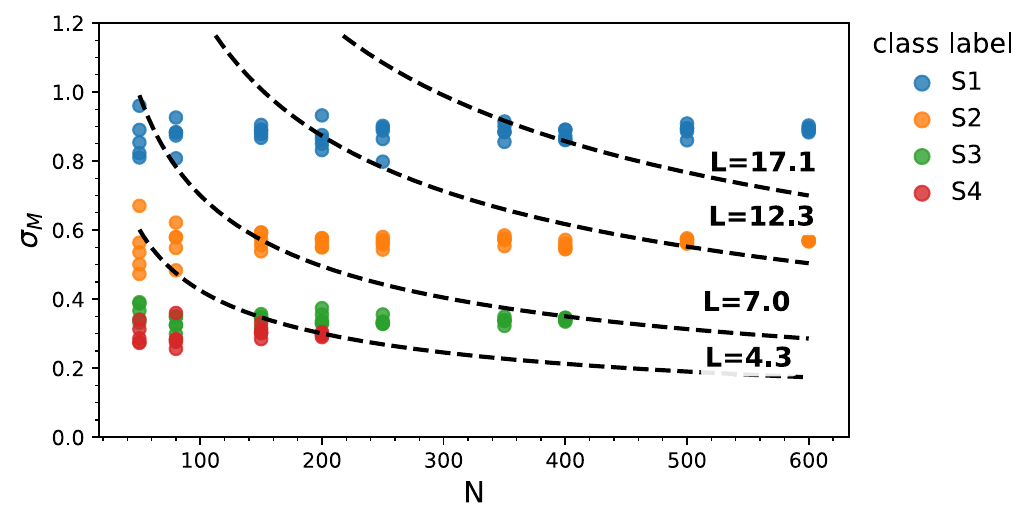}
  \caption{Survey design space:  the standard deviation in
  planetary mass, $\sigma_M$, versus sample size, $N$. Each point represents one survey, with colours
  indicating mass class (S1--S4). Contours show curves of constant mass
  leverage, $L_{\mathrm{mass}} \propto \sqrt{N}\,\sigma_M$. As expected, S1 surveys span the
  widest mass range and achieve the highest leverage at every $N$.}
  \label{fig:design}
  \end{figure}

\subsection{Survey Leverage}\label{sec:leverage}
Survey leverage along a single axis of
  diversity, $x$, is defined by \citet{Cowan2025} as the quadrature sum of deviations around the sample mean, $\bar{x}$:
      \begin{equation}
          L = \sqrt{\sum_{i=1}^{N} (x_i - \bar{x})^2}.
      \end{equation}
     Since we consider an axis of diversity with significant uncertainty, stellar metallicity, we introduce \emph{normalized} leverage:
    \begin{equation}\label{eq:leverage}
    L_{p} =
    \sqrt{\sum_{i=1}^{N} \frac{(m_i - \bar{m})^2}{\sigma_{m,i}^2}},
    \qquad
    L_{s} =
    \sqrt{\sum_{i=1}^{N} \frac{(s_i - \bar{s})^2}{\sigma_{s,i}^2}},
    \end{equation}
When the axes of diversity carry non-negligible measurement uncertainty, the relevant
quantity is not the raw spread but the spread measured in units of that uncertainty. We
therefore introduce normalized leverage:
where $\sigma_{m,i}$ and $\sigma_{s,i}$ are the uncertainties in planetary mass and stellar
metallicity for each planet. Because Equation~\eqref{eq:leverage} sums squared
\emph{standardised} deviations, it is controlled by the ratio of population spread to
measurement error rather than by either quantity alone: a covariate can be measured very
precisely and still carry little leverage if the population barely varies along it.

\begingroup
\setlength{\abovecaptionskip}{4pt}
\setlength{\belowcaptionskip}{2pt}
\setlength{\tabcolsep}{4pt}
\centering

\captionof{table}{Dynamic range of the two axes of diversity in the filtered Ariel MCS.
The final column is the median per-object contribution to the normalized
leverage of Equation~\eqref{eq:leverage}.}
\label{tab:dynrange}

\begin{tabular}{lcccc}
\hline
Axis &
\shortstack{Median\\$\sigma$} &
\shortstack{Population\\SD} &
Span &
\shortstack{Median\\$|x_i-\bar{x}|/\sigma_i$} \\
\hline
$\mathrm{[Fe/H]}_\star$ & $0.080$ & $0.201$ & $1.45$ & $1.7$ \\
$\log(M/M_{\rm J})$     & $0.054$ & $0.889$ & $5.10$ & $11.4$ \\
\hline
\end{tabular}

\par\medskip
\endgroup
The distinction between the two axes of diversity is one of dynamic range rather than of
measurement quality. Host-star metallicities are in fact the better-measured of our two
covariates, with a median uncertainty of $0.080$~dex; but because the hosts are drawn from
the solar neighbourhood, the population is less diverse, having a standard deviation in
$\mathrm{[Fe/H]}_\star$ of only $0.201$~dex, giving a median per-object leverage
$|x_i-\bar{x}|/\sigma_i$ of just $1.7$. Planetary mass has a smaller median uncertainty in
dex ($0.054$) but a larger population spread ($0.889$~dex), and hence a per-object
leverage of $11.4$ (Table~\ref{tab:dynrange}). Therefore, planetary mass is the
better-spread axis of diversity, by roughly a factor of seven in per-object signal-to-noise.

Figure~\ref{fig:design} shows the resulting $N$--$\sigma_{M}$ design space,
illustrating how survey class and sample size jointly determine the available
leverage. Moving from S1 to S4 progressively removes the low-mass end of the population
and therefore decreases the mass leverage available to constrain the planetary
mass--metallicity trend.

\subsection{The Hierarchical Model}\label{sec:science_eq}
After generating mock Ariel Tier 2 transit surveys, we retrieve population trends with a hierarchical model. We model the planetary atmospheric metallicity in log space,
\(y_i = \log [{\mathrm{(C+O)/H},i}\)], as a linear function of centered
planetary mass \(m_i\) and centered host-star metallicity \(s_i\):
  \begin{equation}\label{eq:metmodel}
      y_i = \alpha_p + \beta_p\, m_{c,i} + \beta_s\, s_{c,i}
            + \epsilon_{\mathrm{int},i},
  \end{equation}
  where
  \[
      m_{c,i} = \log \frac{M_i}{M_J} - \overline{\log \frac{M_i}{M_J}},
      \qquad
      s_{c,i} = s_{\mathrm{true},i} - \overline{s_{\mathrm{obs}}}.
 \]
  The prior on the intercept $\alpha_p$ is centered on the mean atmospheric
  metallicity of the survey, with uncertainty decreasing with sample size:
  $
  \alpha_p \sim \mathcal{N}(\bar{y},\sigma_y/
  \sqrt{N}\bigr).
  $ The prior on the planetary mass--atmospheric metallicity slope, $\beta_p$ , is centered on zero, reflecting no
  preference for either a positive or negative mass--metallicity trend:
  $
  \beta_p \sim \mathcal{N}(0,\Delta y/\Delta m\bigr).
  $ The prior on the stellar-metallicity slope $\beta_s$ is centered on unity,
  assuming a one-to-one scaling between
  stellar and planetary metallicity:
  $
  \beta_s \sim \mathcal{N}(1,\Delta y/\Delta s\bigr).
  $ This reflects the expectation that planets forming in metal-rich
  disks have access to a larger reservoir of heavy elements, which may
  increase their atmospheric enrichment, while allowing
  data to determine the strength of the stellar-metallicity dependence
  \citep{Guillot_2006,Thorngren_2016}. The intrinsic-scatter scale $\varepsilon$ follows a Half-normal prior,
$
\varepsilon \sim \mathrm{HalfNormal}(\sigma_y),
$
which controls the amplitude of the planet-level intrinsic scatter in atmospheric metallicity:
$
\epsilon_{\mathrm{int},i} \sim \mathcal{N}(0,\varepsilon).
$
   Here, $\bar{y}$ and $\sigma_y$ are the mean and standard
  deviation of $y$ within a survey, while $\Delta y$, $\Delta
  m$, and $\Delta s$ are the corresponding spans (max minus min).
  Host-star metallicity has limited dynamic range and sizeable measurement uncertainty, so
  treating the observed value $s_{\mathrm{obs},i}$ as exact would introduce
  errors-in-variables bias. We follow the standard Bayesian treatment for uncertain covariates  \citep{Kelly_2007,hogg2010}: the measured stellar metallicity is
  treated as a noisy observation of an underlying latent value, rather than
  as the predictor itself. This matters because ignoring measurement error biases slope estimates
  toward zero. The response uncertainty and
  intrinsic astrophysical scatter are handled separately in the likelihood,
  so the model distinguishes measurement noise from population-level
  planet-to-planet diversity. We introduce a planet-by-planet latent
  variable, $s_{\mathrm{true},i}$, representing the unobserved true
  stellar metallicity:
  $
        s_{\mathrm{true},i} \sim
        \mathcal{N}\bigl(s_{\mathrm{obs},i},
        \sigma_{s,\mathrm{meas},i}\bigr).
  $
  The MCS reports asymmetric upper and lower uncertainties, which we
  approximate with a symmetric Gaussian error scale,
  $
        \sigma_{s,\mathrm{meas},i}
        =
        \frac{1}{2}
        \left(|e_{\mathrm{lo},i}| + |e_{\mathrm{hi},i}|\right).
  $
  The centered stellar predictor is then
  $
        s_{c,i}
        =
        s_{\mathrm{true},i}
        -
        \overline{s_{\mathrm{obs}}},
        \text{where }  
        \overline{s_{\mathrm{obs}}}
        =
        \frac{1}{N}\sum_i s_{\mathrm{obs},i}.
  $
  Thus, the model propagates stellar-metallicity measurement uncertainty
  through the posterior while keeping the intercept anchored to the
  observed survey centroid.

  The likelihood for the observed atmospheric metallicity is
  \begin{equation}\label{eq:likelihood}
      y_i \sim \mathcal{N}\bigl(\mu_i,\sigma_{\mathrm{obs},i}\bigr),
      \qquad
      \mu_i = \alpha_p + \beta_p\, m_{c,i} + \beta_s\, s_{c,i}.
  \end{equation}
  The total scatter entering the likelihood is the quadrature sum of measurement noise
  and intrinsic astrophysical scatter:
  \begin{equation}\label{eq:obssigma}
      \sigma_{\mathrm{obs},i}
      = \sqrt{\sigma_{y,\mathrm{meas},i}^{2} + \varepsilon^2}.
  \end{equation}
  As $\varepsilon \to 0$, the likelihood is dominated by measurement
  uncertainty alone; as $\varepsilon \gg \sigma_{y,\mathrm{meas},i}$,
  intrinsic astrophysical scatter dominates.

 In the baseline implementation, planetary mass enters the
  model through $\log M_i$, without explicitly propagating
  mass uncertainties. This pragmatic simplification is
  appropriate for the confirmed planets considered here,
  where typical uncertainties in planetary mass  are much smaller
  than the population-level variance in planetary mass: across our sample the mean squared uncertainty in $\log(M/M_{\rm J})$ is
$\langle\sigma_m^2\rangle = 0.024$~dex$^2$ against a population variance of $0.791$~dex$^2$, so the
attenuation factor $\lambda_m = [1 + \langle\sigma_m^2\rangle/\mathrm{Var}(\log M)]^{-1} = 0.971$
\citep{Fuller1987,Carroll2006} and treating masses as exact biases $\beta_p$ low by only $2.9$ per
cent, whereas doing the same would lead to a bias of $24$ per cent for $\mathrm{[Fe/H]}_\star$.

\subsection{Model Comparison and Inference }\label{sec:variants}

  We compare two model variants. The baseline \emph{3D Model} is denoted by Equation ~\ref{eq:metmodel}
  while the reduced \emph{2D Model} is
  \begin{equation}
      y = \alpha_p + \beta_p m_c + \epsilon_{\mathrm{int}}.
  \end{equation}
 In the 2D Model, the dependence of atmospheric metallicity on stellar metallicity is not
  modelled explicitly and therefore contributes to the astrophysical
  variance, inflating the inferred intrinsic scatter scale $\varepsilon$
  relative to the 3D Model. In other words, when a real stellar trend is
  present, the omitted term $\beta_s s_c$ is absorbed into the
  residual scatter. In this framework,
  $\varepsilon$ absorbs both genuine planet-to-planet diversity at fixed
  mass and stellar metallicity, and any unmodelled dependence on other
  axes of diversity, such as equilibrium temperature, irradiation history,
  orbital separation, age, cloud properties, C/O ratio, or formation
  location. Thus, a large inferred $\varepsilon$ may indicate either true
  astrophysical randomness or an incomplete population model that needs to factor in an extra axis of diversity. In this
  sense, intrinsic scatter is also a diagnostic: if adding a new physical
  predictor reduces $\varepsilon$, that predictor explains part of the
  population-level diversity. Thus, comparing the 2D and 3D Models tests whether the
  framework can distinguish a genuine stellar--planetary
  metallicity correlation from intrinsic astrophysical
  scatter.

We sample the posterior using the No-U-Turn Sampler \citep[NUTS;][]{HoffmanGelman2014} as
implemented in NumPyro \citep{Phan2019}. NUTS is a variant of Hamiltonian Monte Carlo, in which the sampler uses
the gradient of the log posterior to follow the geometry of the target distribution, proposing
distant moves that are nonetheless accepted with high probability; the no-U-turn criterion then
sets the length of each trajectory automatically, removing the principal hand-tuned parameter of
Hamiltonian Monte Carlo. We adopt it here because the sampled space is considerably larger than
the four population parameters suggest: the errors-in-variables treatment of
Section~\ref{sec:science_eq} introduces one latent stellar metallicity per planet, so a survey of $N$
planets is sampled in $N+4$ dimensions, up to $604$ for our largest surveys. Gradient-based
samplers remain efficient at this dimensionality, whereas the random-walk behaviour of
Metropolis- or Gibbs-type samplers would mix prohibitively slowly. Gradients are obtained by
exact reverse-mode automatic differentiation of the model, with no finite-difference or analytic
approximation.

For each fit we run four chains, each initialised independently by drawing from a uniform
distribution on $[-2,2]$ in the unconstrained parameter space. During warmup the step size is
tuned by dual averaging to a target acceptance probability of $0.9$ and a diagonal mass matrix is
adapted using expanding windows \citep{Betancourt2017}; the maximum tree depth is $10$. We use
$800$ warmup steps and $800$ posterior draws per chain. This length was chosen empirically, as
the shortest run that converged on the hardest surveys in our grid (those with the smallest
$N$ and lowest leverage, where $\varepsilon$ and $\alpha_p$ are least well constrained) rather
than projected from the number of free parameters. Convergence was assessed with the
rank-normalised split-$\hat{R}$ statistic and the bulk effective sample size (ESS) of
\citet{Vehtari2021}: across our fits, $\hat{R}$ never exceeds $1.006$ and the
minimum ESS across all parameters is $1810$ of $3200$ post-warmup draws. Both are comfortably inside the recommended
thresholds of $\hat{R} < 1.01$ and ESS $> 400$, so longer chains were unnecessary; the total cost
of the survey grid was the binding constraint.

Each of the $180$ mock surveys is fit under both model variants and repeated across three
independent MCMC seeds, yielding a total of $1080$ MCMC fits. Posterior summaries are reported as
the mean, standard deviation, and 16th/84th percentile credible intervals for the intercept,
slopes, and intrinsic scatter scale: $\{\alpha_p,\beta_p,\beta_s,\varepsilon\}$.
  \begin{table*}                                                                
  \centering
\caption{Per-parameter summary statistics for $z_\theta$ across all
surveys.}
\label{tab:zsummary}
\footnotesize
\setlength{\tabcolsep}{4pt}
\renewcommand{\arraystretch}{1.15}
\begin{tabular}{p{0.10\textwidth}p{0.32\textwidth}cccrrr}
  \toprule
  Parameter &
  Description &
  mean($z$) &
  SD($z$) &
  median($z$) &
  max$|z|$ &
  $|z|<1$ &
  $|z|<2$ \\
  \midrule

  $\alpha_p$ &
  Average atmospheric metallicity across the population of planets. &
  $-0.003$ & $1.093$ & $-0.034$ & $2.59$ & $65.5\%$ & $91.0\%$ \\

  $\beta_p$ &
  Slope of atmospheric metallicity with planetary mass,
  $\log(M/M_J)$. &
  $0.099$ & $0.739$ & $0.021$ & $2.37$ & $81.4\%$ & $99.3\%$ \\

  $\beta_s$ &
  Slope of atmospheric metallicity with host-star metallicity. &
  $0.047$ & $0.731$ & $0.027$ & $1.94$ & $82.1\%$ & $100.0\%$ \\

  $\varepsilon$ &
  Intrinsic astrophysical scatter in atmospheric metallicity. &
  $0.003$ & $0.853$ & $0.051$ & $3.36$ & $80.0\%$ & $97.2\%$ \\

  \bottomrule
\end{tabular}                                                                              
      \vspace{0.4em}
      \begin{minipage}{0.95\textwidth}
      \scriptsize                                                               
      \textit{Note.} For each parameter $\theta$, $z_\theta$ is computed
      separately for every mock survey using Equation~\ref{eq:zscore}.          
       The slope and scatter parameters use
      the common full-MCS reference. The mean, standard deviation, and median of the   
      z-scores, as well as the largest absolute z-score, max$|z|$,             
      across the survey ensemble are shown. The columns $|z|<1$ and            
      $|z|<2$ report the percentage of surveys whose z-scores fall              
      within one and two posterior standard deviations of the reference       
      value, respectively.                                                      
      \end{minipage}                                                          
    \end{table*}  

\begin{figure*}
  \centering

  \begin{minipage}{0.95\textwidth}
  \centering
  \includegraphics[width=\textwidth,alt={Posterior reconstruction plots for representative HERMES survey fits.
  The panels compare atmospheric abundance against centred planetary mass
  and centred stellar metallicity, showing data with error bars, posterior
  median trends, and 68 percent credible bands.}]{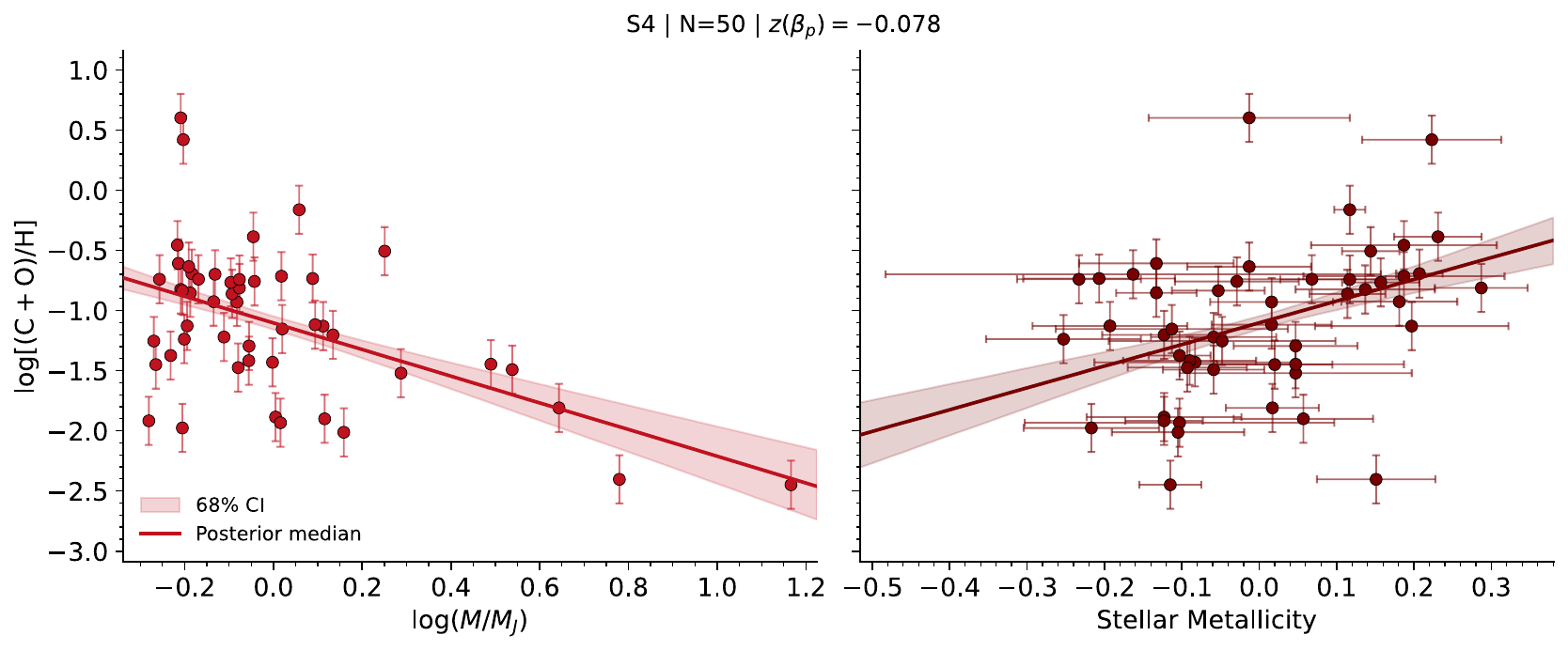}
  \end{minipage}
  \hfill
  \caption{
  The left panel shows planetary atmospheric abundance $\log [{\mathrm{(C+O)/H}}]$ versus
  centered $\log(\frac{M}{M_J})$, and the right panel gives $\log [{\mathrm{(C+O)/H}}]$ versus centered stellar
metallicity $[\mathrm{Fe/H}]_\star$. In both cases, the
posterior median trend and its 68\% credible band are shown. The
error bars correspond to $0.2$ dex for $\log [{\mathrm{(C+O)/H}}]$ and true measurement uncertainty on stellar metallicity from the MCS.}
  \label{fig:survey_fits}
  \end{figure*}

\section{Results}
\subsection{Hierarchical Model Survey Fits}\label{sec:fits}

We generate mock surveys as defined in Section~\ref{sec:classes}
and run the 3D HERMES model on each survey. Figure~\ref{fig:survey_fits}
shows posterior reconstructions for an example survey in the S4 mass class. As expected, S1 class surveys have the highest leverage and the tightest credible
bands, while S4 have the lowest leverage and the widest
bands. All surveys successfully recover the injected trends,
showing that HERMES performs well across the full range of survey
geometries. 

\subsubsection{Z-Score Calibration}\label{sec:zscores}
  To assess posterior calibration, we define a $z$ score for each               
  parameter $\theta \in \{\alpha_p, \beta_p, \beta_s, \varepsilon\}$ and
  survey $k$ as                                                                 
  \begin{equation}\label{eq:zscore}                                             
      z_\theta^{(k)}                                                          
      = \frac{\hat{\theta}^{(k)} - \theta_{\mathrm{true}}^{(k)}}                 
             {\sigma_\theta^{(k)}},
  \end{equation}                                                                
  where $\hat{\theta}^{(k)}$ is the posterior mean,                             
  $\sigma_\theta^{(k)}$ is the posterior standard deviation, and              
  $\theta_{\mathrm{true}}^{(k)}$ is the ``true'' value per survey $k$,          
  taken from the population-level fit obtained on the full MCS. For the slope and scatter                   
  parameters, this reference is the same for every survey. The                  
  intercept $\alpha_p$ requires a survey-specific reference because it        
  represents the average atmospheric metallicity at the mean planetary       mass and mean stellar metallicity of a given survey; since the S2--S4       
  classes progressively remove the low-mass planets, their covariate          
  means differ substantially from the full catalogue, and the intercept         
  is expected to shift between mass classes rather than being recovered
  around a single value. We therefore project the population-level fit       
  onto the covariate means of each survey to obtain                             
  $\alpha_{p,\mathrm{true}}^{(k)}$, so that the same calibration 
  applies across all mass classes. If the posteriors are well                   
  calibrated, all resulting z-scores should approximately follow a  Gaussian distribution, with $|z|<1$ for roughly 68\% of surveys and           
  $|z|<2$ for roughly 95\% of surveys. Table~\ref{tab:zsummary} is built by computing $z_\theta^{(k)}$ from Equation~\eqref{eq:zscore}
for every mock survey and pooling across all four mass classes and the full range of sample sizes
and leverages. The reference value each survey is compared against is obtained by fitting the
model to that survey's entire mass class. This is necessary because the mock catalogue was generated once, with the intrinsic scatter following a single distribution: the slope present in the data is therefore not exactly the injected value, and differs slightly between the nested classes, so comparing every class to one
full-catalogue value would add calibration error.
Table~\ref{tab:zsummary} summarizes the per-parameter model                   
  performance, including the $z$-score statistics, across all surveys.        
  The slope parameters, $\beta_p$ and $\beta_s$, the intrinsic-scatter      
  scale, $\varepsilon$, and the intercept, $\alpha_{p}$, are all well calibrated: their $z$-score standard       
  deviation, $\mathrm{SD}(z)$, is close to unity, and their coverage
  fractions are near the expected Gaussian values.

\subsection{Posterior Uncertainty versus Leverage and Survey size}\label{sec:expected}

  For ideal linear regression, the uncertainty on a slope parameter is
  expected to scale inversely with the survey leverage along the that specific dimension of diversity \citep{Cowan2025},
 \[
      \sigma_{\beta} \propto \frac{1}{L}.
 \]
  In practice, this idealized scaling is modified by intrinsic
  astrophysical scatter and measurement uncertainty. To quantify the empirical scaling in the simulations, we fit
  power-law relations of the form
 \[
      \sigma_\theta = a\,L_{x}^{\gamma},
\]
  and report the fitted exponent $\gamma$ and its uncertainty for each axis of diversity, $x$. By testing whether $\gamma$ is closer to -1, we are able to
assess whether leverage remains the dominant control on slope precision once the problem is extended from a single
  predictor to a multidimensional hierarchical model.

\begin{figure*}
  \centering
  \includegraphics[width=0.85\textwidth, alt={Comparison of posterior uncertainty in the planetary mass
  slope against mass leverage and stellar-metallicity leverage for surveys with
  N equals 50. Uncertainty decreases most clearly with mass
  leverage and is lower overall for the larger sample size.}
  ]{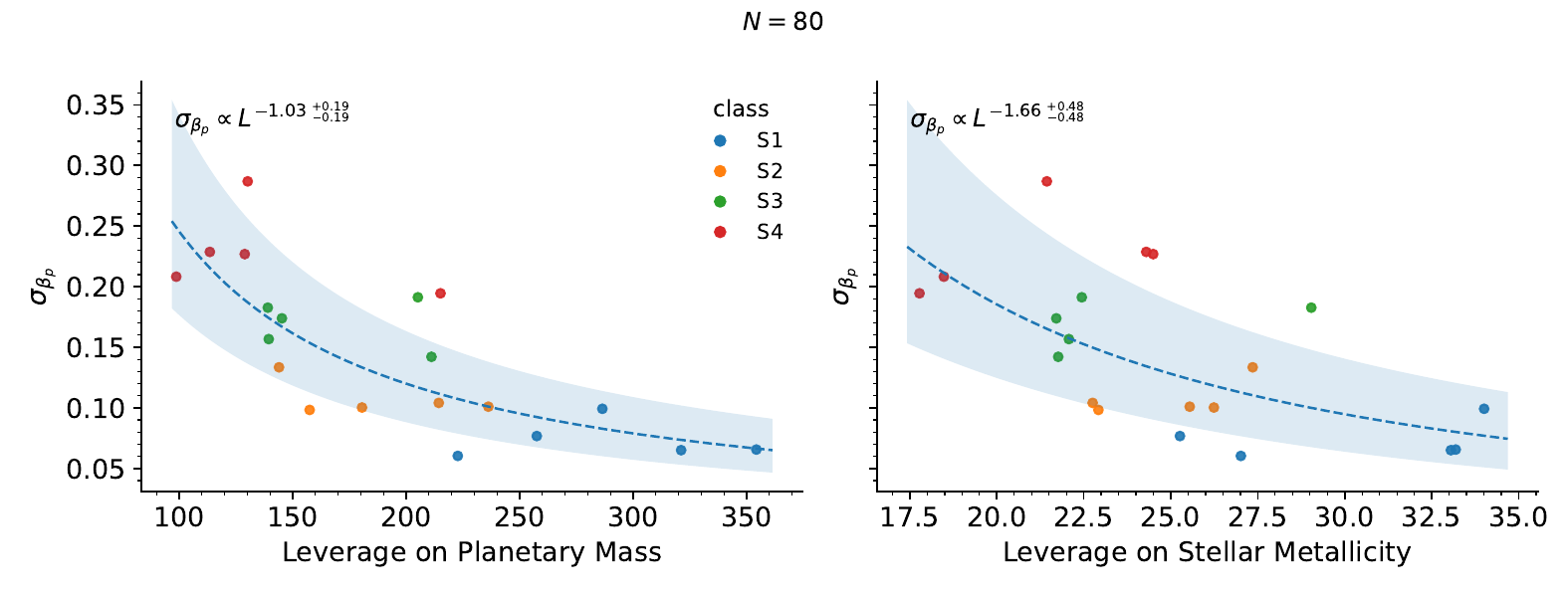}

  \vspace{0.3em}

  \includegraphics[width=0.85\textwidth, alt={Comparison of posterior uncertainty in the planetary mass
  slope against mass leverage and stellar-metallicity leverage for surveys with
  N equals 250. Uncertainty decreases most clearly with mass
  leverage and is lower overall for the larger sample size.}]{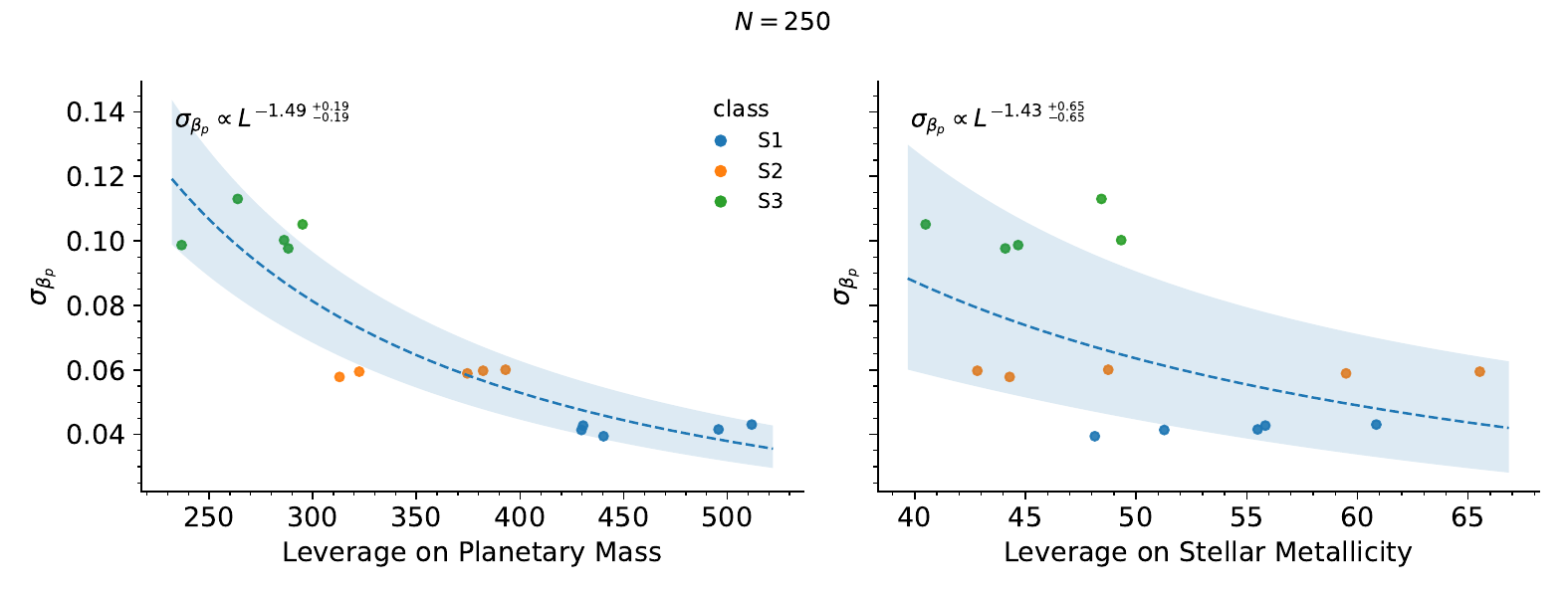}
  
 \vspace{0.3em}

 \caption{Posterior uncertainty in the mass--metallicity slope as a function of
  planetary-mass leverage (left panels) and host-star-metallicity leverage
  (right panels), shown for surveys with 80 and 250 planets. Each point
  represents one mock survey. Dashed curves show power-law fits, with the
  shaded regions indicating the corresponding prediction bands. As sample
  size increases, the fit approaches the inverse-leverage
  expectation and the overall uncertainty decreases, while the qualitative
  dependence on survey leverage remains the same. Since the leverage
  values are normalized by their measurement uncertainties
  (Equation~\ref{eq:leverage}), the planetary-mass leverage is generally
  larger than the host-star-metallicity leverage. Moreover, Leverage increases as $\sqrt{N}$, between the top and bottom panels, roughly a factor of two.}
  \label{fig:lev_bp}
  \end{figure*}

  Figure~\ref{fig:lev_bp} shows $\sigma_{\beta_p}$ as a function of both
  $L_{p}$ and $L_{s}$ for surveys with $N=80$ and $N=250$ planets.
  As expected, $L_{p}$ is the more reliable predictor of the
  mass-slope precision: surveys with larger mass leverage yield smaller
  posterior uncertainty on $\beta_p$. This confirms that the precision of
  the planetary mass--metallicity slope is governed primarily by the
  lever arm in planetary mass. The scaling relation is consistent across
  sample sizes: increasing $N$ lowers the overall uncertainty floor, but
  preserves the leverage dependence. 

At fixed sample size, $\sigma_{\varepsilon}$ and $\alpha_{p}$ show little dependence on either
  leverage axis: the posterior constraints on these parameters improve simply as $\sqrt{N}$, as shown in Figure~\ref{fig:unc_vs_N}. The slope parameters
  respond
  most strongly along their relevant axes of diversity: the planetary
  mass--atmospheric metallicity slope, $\beta_p$, is close to the inverse leverage
  expectation, while the stellar--atmospheric metallicity slope, $\beta_s$,
  also improves with stellar-metallicity information, although more weakly. This weaker scaling likely reflects the limited spread in $L_s$, as
  shown in Figure~\ref{fig:lev_bp}. This is driven by the narrower
  range of $[\mathrm{Fe/H}]_\star$ and its significant measurement
  uncertainty. In
  contrast,
  the intercept, $\alpha_p$, and intrinsic scatter, $\varepsilon$, improve
  mainly
  because larger surveys reduce the overall posterior uncertainty (see Figure~\ref{fig:unc_vs_N}). Thus, a survey with more planets improves constraints on
  all parameters along multiple axes of diversity, while leverage along the relevant physical axis most
  efficiently
  constrains the corresponding slope.

\begin{figure*} 
\centering
\includegraphics[width=0.95\textwidth,  alt={Heatmap showing the fraction of mock surveys where the 3D model is
  favoured over the 2D model by WAIC as a function of intrinsic scatter and
  sample size. Recovery is highest for large samples and lowest for small
  samples at high intrinsic scatter.}]{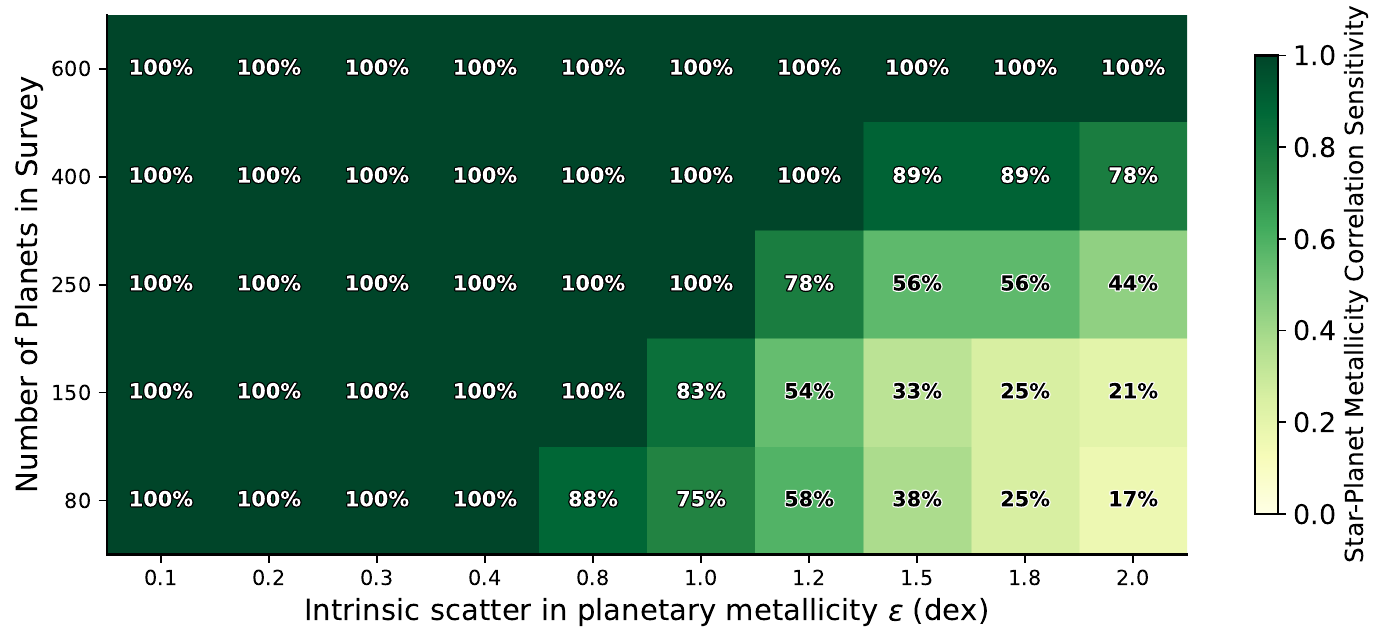}
\caption{The heatmap shows the practical recovery threshold
for the stellar--planetary metallicity correlation: small and moderate
surveys lose sensitivity once intrinsic scatter becomes large, whereas
Ariel-scale samples maintain high recovery fractions across the tested
scatter range.}
\label{fig:scatter_heatmap}
\end{figure*}
\section{Recovering the Stellar--Planetary Metallicity Correlation}\label{sec:stellar_det}

As discussed in Section~\ref{sec:variants}, the 3D Model explicitly includes the stellar metallicity slope $\beta_s$, whereas the 2D Model omits it entirely, absorbing any correlation between stellar and planetary metallicity into the intrinsic scatter $\varepsilon$. Here, we consider how large the intrinsic astrophysical
  scatter can become before the stellar--planetary metallicity correlation
  can no longer be distinguished from unexplained planet-to-planet
  diversity, and whether the added stellar-metallicity term improves predictive accuracy enough to justify the additional model
  complexity. To quantify the fraction of surveys for which the 3D model outperforms the 2D Model, we perform a model comparison using the Watanabe--Akaike Information Criterion (WAIC).

\subsection{The Watanabe--Akaike Information Criterion}\label{sec:waic}

To determine whether including the stellar metallicity term provides
a meaningful improvement in predictive accuracy, we compare the 3D
and 2D Models (Section~\ref{sec:variants}) using the Watanabe--Akaike Information Criterion \citep{watanabe2010}. This serves an analogous role to the Akaike and Bayesian Information
  Criteria (AIC/BIC) but is better suited to hierarchical models, where         
  the effective number of parameters is not fixed but emerges from the          
  posterior and the data 
\citep{gelman2014understanding}. WAIC
estimates the expected log pointwise predictive density (elpd), a
measure of predictive
accuracy. This is the quantity a model
would score if it were used to predict new data: for each observation we ask how much posterior
predictive probability the fitted model assigns to a hypothetical replacement measurement at that
point. It therefore
measures out-of-sample predictive accuracy rather than goodness of fit, and penalises a model that
reproduces the observed data only by over-fitting it. Since  'new observations' are not available, WAIC
estimates the elpd from the in-sample pointwise log-likelihood draws, corrected by
the posterior variance of the pointwise log-likelihood,

For each survey we compute:
\begin{equation}\label{eq:dwaic_stellar}
    \Delta\mathrm{elpd}_{\mathrm{WAIC}}
        = \mathrm{elpd}_{\text{3D Model}}
         - \mathrm{elpd}_{\text{2D Model}}.
\end{equation}
A positive $\Delta\mathrm{elpd}_{\mathrm{WAIC}}$ indicates that the 3D mass--stellar--atmospheric metallicity model outperforms the 2D mass--atmospheric metallicity model, and thus that the framework has successfully recovered the stellar signal as distinct from intrinsic scatter.

\subsection{Recovering the Stellar--Planetary Metallicity Correlation}\label{sec:science_q}

For two planets of similar mass, orbiting the same type of star, we may not observe the same planetary metallicity due to intrinsic astrophysical scatter \citep{Turrini_2023}.  As we enter the era of comparative exoplanetary climatology, factoring intrinsic astrophysical scatter in survey design becomes an important factor, especially for Ariel.
The central question becomes: \emph{given increasing intrinsic
astrophysical scatter, for what fraction of surveys can we recover the
correlation between stellar and planetary metallicity using
hierarchical modelling?} We probe this by systematically increasing the injected intrinsic scatter across surveys of different sample sizes. For each survey,
  we record whether the 3D Model is favoured over the 2D Model according
  to the WAIC comparison defined in Section~\ref{sec:waic}.

 Figure~\ref{fig:scatter_heatmap} shows the fraction of mock surveys for
  which WAIC favours the 3D Model over the 2D Model as a function of
  intrinsic scatter and sample size. This result sets a clear detection threshold for the stellar--planetary metallicity correlation. At low intrinsic scatter, representing deterministic planet formation and evolution, the correlation is recovered for nearly all surveys. As $\varepsilon$ increases, recovery first fails for the smallest samples while
  larger surveys retain sensitivity to the star-planet metallicity term over a
  wider range of intrinsic scatter. Surveys with $N \leq 150$ show a clear
  drop in sensitivity once $\varepsilon \gtrsim 0.8$--$1.0$
  dex, whereas Ariel-scale surveys remain robust to substantially larger
  scatter. In particular, the $N=600$ surveys favour the 3D Model in 100\%
  of cases across the full scatter range tested. The median
  $\Delta\mathrm{elpd}_{\mathrm{WAIC}}$ follows the same trend, declining
  approximately as $\varepsilon^{-2}$.

\section{Summary and Discussion}\label{sec:summary}

 We extended the survey leverage framework of \citet{Cowan2025} by
  propagating measurement uncertainty for $[\mathrm{Fe/H}]_\star$ into the
  posterior and avoiding the attenuation bias that would arise from
  treating observed stellar metallicities as exact (Section~\ref{sec:science_eq}). We also explicitly modelled intrinsic astrophysical scatter as a free parameter $\varepsilon$, quantifying the stochastic nature of planetary formation, evolution and any other unmodelled axes of diversity. The $z$-score calibration (Table~\ref{tab:zsummary}) confirmed that the resulting posteriors for $\beta_p$ and $\beta_s$ are well-calibrated ($\mathrm{SD}(z) \approx 1$), validating that the framework is neither overconfident nor overly conservative. 

  Section~\ref{sec:science_q}       
  demonstrated that the ability to distinguish
  the stellar--planetary metallicity correlation from intrinsic            
  astrophysical scatter depends jointly on $N$ and $\varepsilon$, with        
  survey size becoming the decisive factor once scatter exceeds           
  $\sim\!0.8$\,dex (Figure~\ref{fig:scatter_heatmap}). Below this               
  threshold, even moderate surveys ($N \leq 250$) achieved near-complete      
  recovery; above it, only Ariel-scale samples ($N \geq 400$) maintained      
  high detection fractions.

\subsection{Implications for Ariel Target Selection}
 
The story is not as simple as ``maximize leverage'', but to optimize both
  leverage and survey size, for example through simulated annealing
  \citep{panek2026}, while also improving constraints on $[\mathrm{Fe/H}]_\star$, stellar age, and other harder-to-measure axes of diversity. Different parameters respond to different design axes. A survey optimized purely for mass leverage could sacrifice stellar-metallicity diversity and leave $\beta_s$ poorly constrained. Ariel's target selection must therefore balance leverage across multiple axes of diversity if the goal is to constrain the full multidimensional mass--metallicity relation.  
  
Sample size $N$ plays an important but distinct role to survey leverage. Increasing $N$ lowers the overall uncertainty for all parameters (Figure~\ref{fig:unc_vs_N}), but the \emph{relative} importance of leverage versus $N$ differs by parameter. For $\beta_p$, increasing $L_{p}$ at fixed $N$ is more efficient than increasing $N$ at fixed mass range. For the intercept $\alpha_{p}$ and intrinsic scatter $\varepsilon$, which are largely insensitive to
  leverage at fixed $N$, sample size is the primary control parameter.
              
  This interplay between sample size and astrophysical scatter has direct implications        
  for Ariel's tiered observing strategy.                                      
  \citet{Radica2026}                         
  showed that Tier~1 observations (the lowest-precision tier in the          
  Ariel survey design) already yield $\lesssim 1.5$\,dex constraints on        
  H$_2$O and CO$_2$ for giant planets, irrespective of cloud                    
  conditions. Our              
  results show that the $\sim\!1000$-planet Tier~1 sample can compensate        
  through sheer statistical power: Figure~\ref{fig:scatter_heatmap}             
  confirms that $N = 600$ surveys recover the stellar correlation at            
  100\% even for scatter as large as 2.0\,dex. Larger observational uncertainties impact the problem equivalently to larger astrophysical variance. In other words, the            
  breadth of the Tier~1 survey may be more valuable for                         
  population-level trend recovery than the depth of a smaller Tier~2            
  survey, precisely because the additional planets contribute                
  leverage on multiple axes simultaneously. A large, lower-precision Tier 1 survey can still outperform a
     smaller high-precision survey for detecting population trends if the trend
     is broad and the survey size spans enough diversity.                                    
  This conclusion extends the single-axis leverage framework of                 
  \citet{Cowan2025} into a genuinely realistic multidimensional setting. In one            
  dimension, a survey designer faces a clean trade-off: a small,                
  high-leverage sample can match the slope precision of a larger but            
  less diverse one. 

  In multiple dimensions, however, the survey must            
  achieve adequate leverage on \emph{every} axis of diversity                   
  simultaneously. Optimizing           
  for one does not guarantee the other. As \citet{panek2026}                    
  demonstrated, automated target selection strategies that balance              
  leverage across multiple axes of diversity tend to converge toward            
  large, broadly representative samples---close to the maximal survey size. The practical implication is that Ariel target selection should not treat
  survey tier, leverage, and sample size as independent choices.  HERMES therefore builds on the
  leverage framework of \citet{Cowan2025}, and on the
  optimization strategy of \citet{panek2026}, by showing that multidimensional
  population inference naturally pushes survey design toward large, diverse
  samples that remain close to maximal $N$ while preserving leverage across the
  relevant physical axes. This result is encouraging for Ariel: even if the true astrophysical scatter exceeds current estimates, Ariel's
  planned sample size should provide enough statistical power to distinguish
  the stellar-metallicity signal from intrinsic astrophysical scatter. Conversely, smaller precursor surveys may struggle to distinguish the two unless they target populations with intrinsically low scatter or achieve unusually high leverage in stellar metallicity. These results further emphasize Ariel's need for precise, homogeneous
  stellar characterization. Reducing uncertainties in
  $[\mathrm{Fe/H}]_\star$ would increase stellar-metallicity leverage and
  make it easier to separate a genuine stellar contribution from intrinsic
  planet-to-planet scatter.

\subsubsection{The cost of leverage}
\label{sec:cost}

All surveys are ultimately limited by observing time, and it is natural to ask whether the
high-leverage samples are affordable.
Mass leverage is relatively cheap to obtain in a transit spectroscopy survey, however, for this
target sample the two are aligned rather than opposed. The reason is a selection effect in
the candidate sample itself: the low-mass end is populated preferentially by planets
transiting small, bright, nearby M and K dwarfs, whose transit depths per atmospheric scale
height are large, while the high-mass end is dominated by dense, compact companions around
fainter FGK hosts with correspondingly small atmospheric signals.

This conclusion is not specific to Ariel. HERMES takes a sample size, a set of leverages
along each axis of diversity, and a per-planet response uncertainty, and returns the
achievable precision on the population trend; the facility enters only through those
quantities. Homogeneous HST/WFC3 transmission surveys \citep{Tsiaras2018} occupy the
small-$N$, moderate-precision corner of this space, and Spitzer phase-curve compilations
\citep{Cowan_2011,dang2025} provide the analogous population sample for thermal
observables. JWST occupies the corner opposite Ariel Tier~1, delivering far higher
per-planet precision on a much smaller and (crucially) inhomogeneously selected
sample, so that its leverage is set by proposal-driven target selection rather than by
survey design. Figure~\ref{fig:scatter_heatmap} can then be read as a general statement about which
regime wins for a given level of intrinsic astrophysical scatter: where scatter dominates
the per-planet measurement error, breadth beats depth, and where it does not, the reverse
holds.

In addressing whether rocky planets around M dwarfs
retain atmospheres, \citet{Ih2026} used a genetic algorithm to select the most informative set of JWST
targets and find that strong population-level evidence is obtainable within roughly $500$
observing hours using a ``wide and shallow'' strategy. Together with this work and automated
target-selection results of \citet{panek2026}, it indicates that optimal survey design for
population-level science converges on large, broadly representative samples across a range
of facilities and science cases.

 \subsection{Outlook for Population-Level Inference}

Although we have focused on the planetary-mass--stellar--atmospheric metallicity
  relation, the framework is not specific to this use case. The same
  hierarchical model can be applied to any population-level trend in
  which a response variable is measured with uncertainty and is expected
  to depend on multiple axes of planetary or stellar diversity. Examples
  include phase-curve observables such as dayside temperature, nightside
  temperature, hotspot offset, or heat-redistribution efficiency as
  functions of irradiation, rotation, gravity, and atmospheric composition
  \citep[e.g.][]{Keating2022,dang2025}. More generally, HERMES can be used either
  \emph{a priori}, to forecast which survey designs will constrain a
  proposed multidimensional trend, or \emph{a posteriori}, to determine
  which physical axes explain the scatter in an observed population.
\section*{Acknowledgements}

WMFN and NBC thank the reviewers for their feedback in improving the manuscript. NBC acknowledges support from a Canada Research Chair and NSERC Discovery Grant. WMFN acknowledges support from the Carl Reinhardt and Kharusi Family Science Fellowships. The authors are also grateful to members of the Ariel Science Team and Ariel Consortium, who helped sharpen these ideas. The authors thank the Trottier Space Institute and l’Institut de recherche sur les exoplanètes for their financial support and dynamic intellectual environment. WMFN would also like to thank the McGill Exoplanet Characterization Alliance for their support; Jennifer Glover for helpful insight and discussion on data visualization; Jayden Ryga, Stephanie Merkl, Sanaa Shaikh, and Dr. Nayyer Raza for their encouragement, support, and advice.

\bsp
\section*{Data Availability}
The data underlying this article will be shared on reasonable request to the corresponding author.

\section*{Conflict of Interest}
Authors declare no conflict of interest.

\bibliographystyle{rasti}
\bibliography{refs}
  \appendix

  \section{Additional Tables and Figures}

\clearpage

  \begin{figure*}
  \centering
  \includegraphics[width=0.9\textwidth,alt={Two-panel plot of alpha-p z-scores against mass leverage and stellar-
  metallicity leverage. Points are coloured by survey class, with a horizontal
  band marking the well-calibrated region between z equals minus one and plus
  one. Centered around a Gaussian. Points are clustered by survey class away from the mean. Only S1 class surveys are within 1 sigma}]{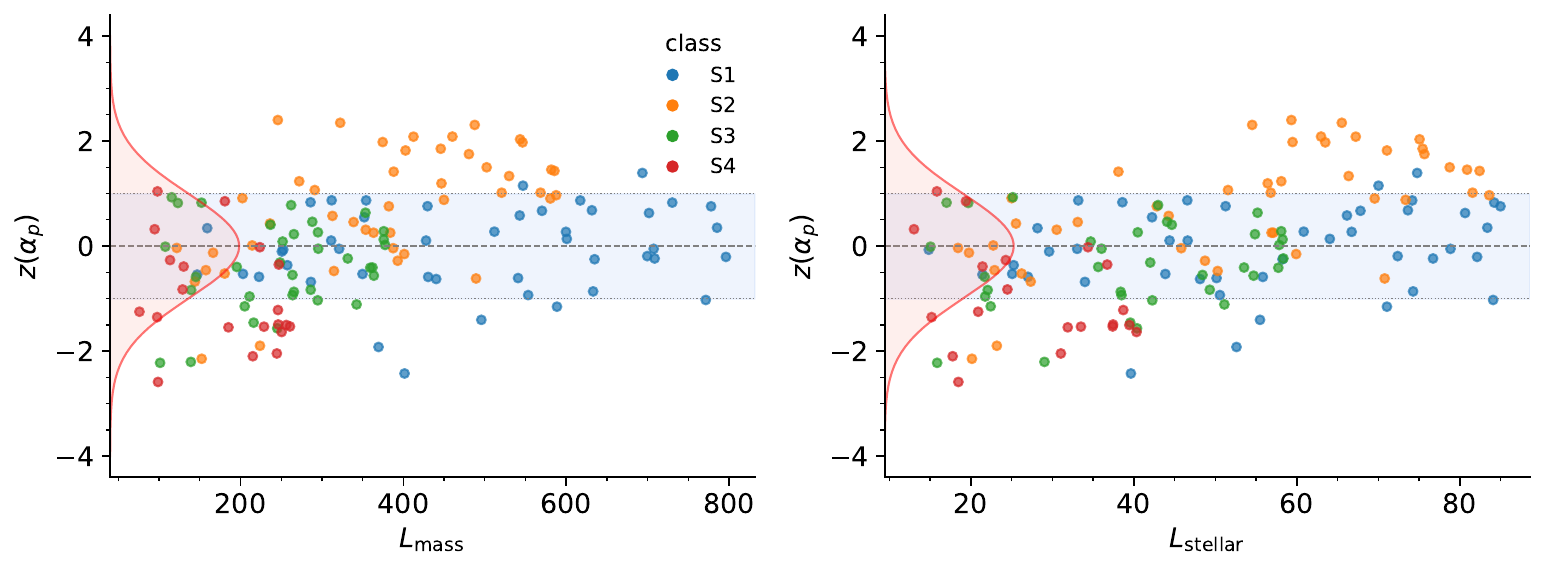}

  \vspace{0.3em}

  \includegraphics[width=0.9\textwidth,alt={Two-panel plot of beta-p z-scores against mass leverage and stellar-
  metallicity leverage. Points are coloured by survey class, with a horizontal
  band marking the well-calibrated region between z equals minus one and plus
  one. Centered around a Gaussian. Most points are within 1 sigma}]{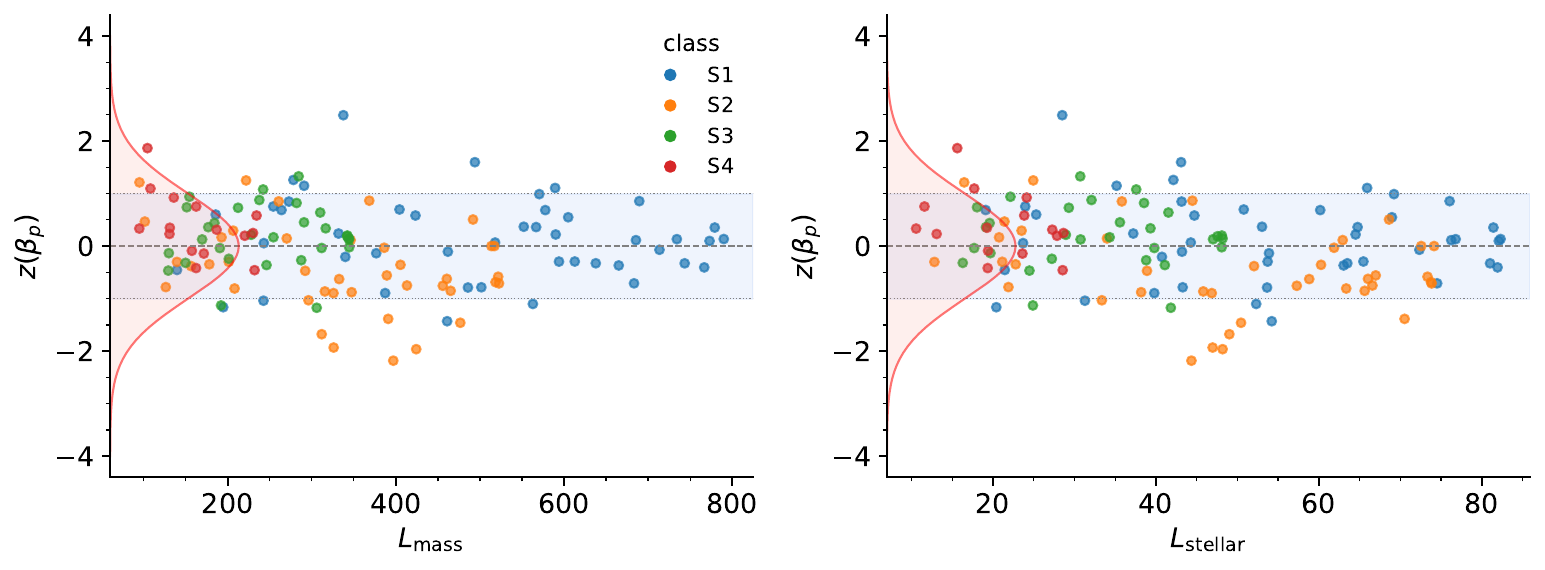}

  \caption{$z(\beta_p)$ as a function of $L_{p}$ (left) and
  $L_{s}$ (right), with points coloured by survey class
  S1--S4. The shaded band marks the nominal well-calibrated region
  $|z|<1$, with the dashed horizontal line indicating $z=0$. As discussed in
  Section~\ref{sec:zscores}, both $z(\beta_p)$ and  $z(\alpha_p)$
  are well calibrated and show supportive evidence for Gaussianity}
  \label{fig:z_p}
  \end{figure*}

  \begin{figure*}
  \centering
  \includegraphics[width=0.85\textwidth, alt={Posterior uncertainty for alpha-p, beta-p, beta-s, and epsilon plotted
  against sample size. Points are coloured by survey class, with fitted power-
  law trends showing that uncertainties generally decrease as sample size
  increases.}]{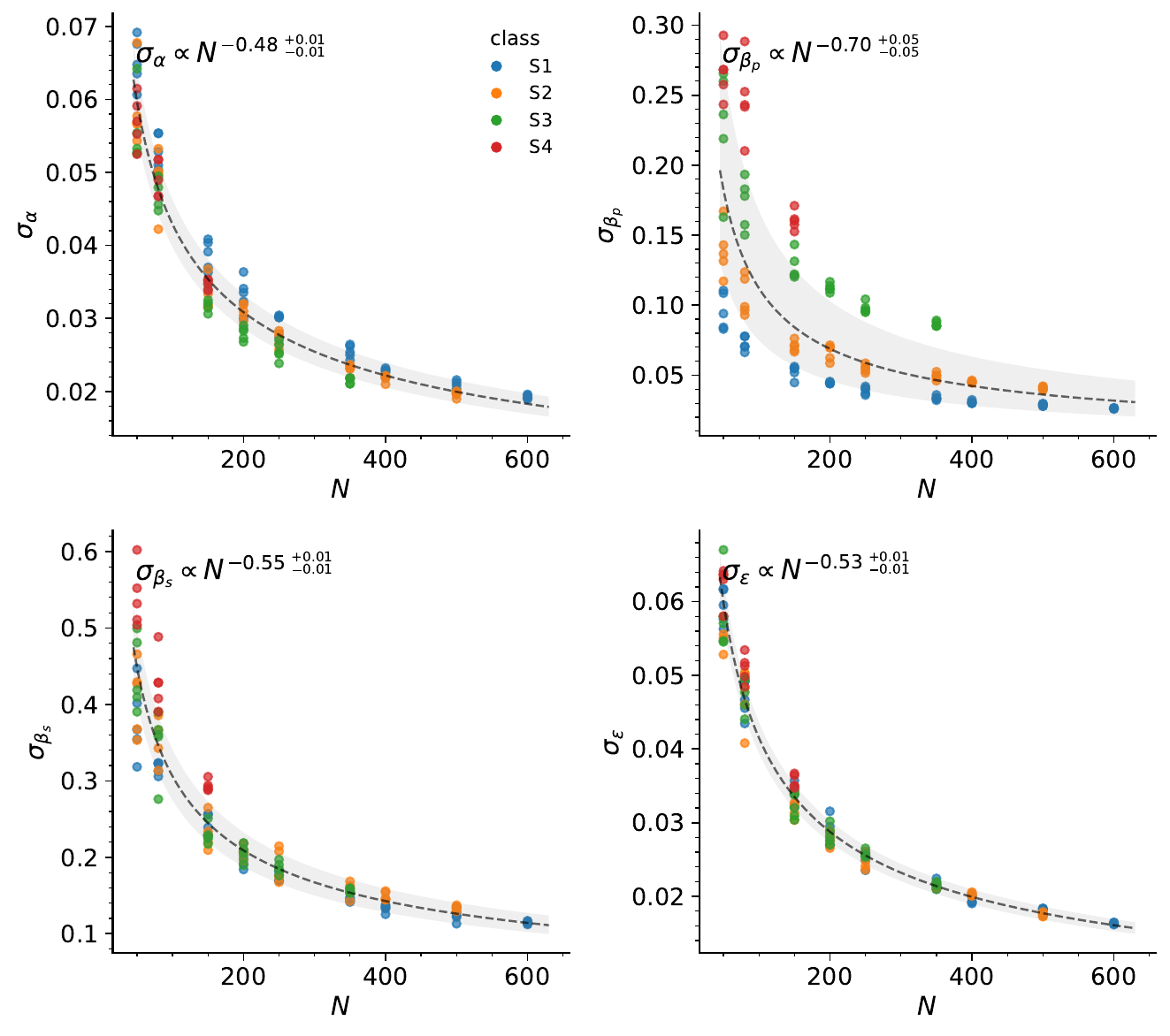}
  \caption{Posterior uncertainty vs.\ $N$ for all four parameters.
  Points = individual surveys colored by class; dashed = power-law fit;
  grey band = prediction interval. Annotation = scaling exponent with
  1-$\sigma$ bounds. The planetary mass--metallicity slope shows more scatter because it mostly scales as $L^{-1}$}
  \label{fig:unc_vs_N}
  \end{figure*}
\label{lastpage}
\end{document}